\documentclass[preprint]{aastex}

\usepackage{amsmath}

\begin{document}

\title{Positions of equilibrium points for dust particles in
the circular restricted three-body problem with radiation}

\author{P. P\'{a}stor}
\affil{Tekov Observatory, Sokolovsk\'{a} 21, 934~01 Levice, Slovak Republic}
\email{pavol.pastor@hvezdarenlevice.sk,~pastor.pavol@gmail.com}

\begin{abstract}
For a body with negligible mass moving in the gravitational field
of a star and one planet in a circular orbit (the circular restricted
three-body problem) five equilibrium points exist and are known as
the Lagrangian points. The positions of the Lagrangian points are
not valid for dust particles because in the derivation of the Lagrangian
points is assumed that no other forces beside the gravitation act on
the body with negligible mass. Here we determined positions
of the equilibrium points for the dust particles in the circular restricted
three-body problem with radiation. The equilibrium points are located
on curves connecting the Lagrangian points in the circular restricted
three-body problem. The equilibrium points for Jupiter are distributed
in large interval of heliocentric distances due to its large mass.
The equilibrium points for the Earth explain a cloud of dust particles
trailing the Earth observed with the {\it Spitzer Space Telescope}.
The dust particles moving in the equilibrium points are distributed
in interplanetary space according to their properties.
\end{abstract}


\section{Introduction}
\label{sec:intro}

Three bodies with masses moving under the action of their mutual
Newtonian gravitation constitute the three-body problem in
celestial mechanics. When one body has negligible mass with respect
to the remaining two, we have the framework of the restricted three-body
problem. The body with negligible mass does not influence the motion
of the remaining two bodies, but its motion is determined by both of them.
If the remaining two bodies orbit around their center of mass
in circles, then the gravitational problem is called
the circular restricted three-body problem (CR3BP). In the CR3BP
the body with negligible mass can remain stationary in a reference
frame rotating with the remaining two bodies around their center
of mass. The stationary location of the body with negligible
mass in the rotating frame is called the equilibrium point.
The body with negligible mass located in the equilibrium point
moves in a circular orbit with a constant speed in an inertial
reference associated with center of mass of the two heavier
bodies. Mutual distances of all three bodies are constant.
Five such equilibrium points exist in the CR3BP.
The points are called the Lagrangian points in honor
of mathematician J. L. Lagrange. Three points are located
on the line determined by two positions of the heavier bodies.
The two remaining points create two equilateral triangles
with the line connecting the two heavier bodies as a common
base. Depending on the mass ratio of the two heavier bodies
the points in the equilateral triangles can be stable. In 1906
first asteroid named Achilles was discovered near the stable Lagrangian
point of Jupiter and the Sun. Today known asteroids near these
stable points form Trojan group.

Unlike larger bodies dust particles can be significantly
affected by weak non-gravitational forces.
The Lorentz force is only important for submicrometer
particles \citep{dohnanyi,leinert,dermott}.
These particles have large ratio of $Q/m$ ($Q$ is the charge
of the dust particle and $m$ is the mass of the dust particle).
The collisions among the dust particles are important
for particles of radii larger than hundred of micrometres,
approximately \citep{grun,dermott}. The motion of micron-sized dust
particles in the inner part of the Solar system is influenced beside
the gravitation mainly by solar electromagnetic radiation and solar wind.
Consequences of the electromagnetic radiation on the motion
of the interplanetary dust particles were already discussed in
\citet{poynting}. \citet{robertson} using relativity theory derived
the equation of motion of a perfectly absorbing and
thermally equilibrated sphere in covariant form.
On the basis of these works influence of the electromagnetic radiation
on the motion of the spherical dust particle is usually called
the Poynting--Roberson (PR) effect. The PR effect contains
a radial radiation pressure term for which inclusion in the CR3BP
is equivalent to inclusion of a less massive star in the CR3BP
and this changes the positions of the Lagrangian equilibrium
points if the radial radiation pressure term is considered.
This was outlined by \citet*{colombo} and \citet{schuerman}.
They also studied the stability of the equilibrium points in
the CR3BP with the PR effect. \citet{murray} systematically
discussed the dynamical effect of general drag in the planar
CR3BP. \citet*{LZJ} investigated the effect
of the PR effect and the radial solar wind in the restricted three-body
problem. However, some of aspects of the motion of the dust particles
in the vicinity of their equilibrium points in the CR3BP with radiation
are still unknown. In this work we use an improved derivation
of the PR effect presented in \citet{icarus}, which followed
results derived in \citet{PRI,klacka2004,MEPR,relation}.
The improved derivation take into account arbitrary optical properties
of particle with material distributed in a spherically symmetric way.
The improved derivation leads to the same result for the case considered
in \citet{robertson}. The relativistic derivation of the PR effect
can be generalized also for the solar wind. This was done in
\citet{covsw}.

\section{Equilibrium points}
\label{sec:equilibrium}

Let us consider the motion of a dust particle in the vicinity of a star with
one planet in a circular orbit. The star produces spherically symmetric
electromagnetic radiation and stellar wind which will be taken
into account. Since orbital speed of the star is slow (small radius
of the orbit and small angular velocity), relativistic corrections caused
by transformations from a frame associated with the center of mass of the star
and the planet into the frame associated with the star can be neglected.
The accelerations of the dust particle caused by the electromagnetic
radiation and the stellar wind will be determined in a reference frame
associated with the star using the theory of relativity to first order
in $v' / c$ ($v'$ is the speed of the dust particle with respect to
the star and $c$ is the speed of light), first order in $u' / c$ ($u'$
is the speed of the stellar wind with respect to the star), and first order
in $v' / u'$. In this case the equation of motion of the dust particle
in the reference frame associated with the star is
\begin{align}\label{star}
      \frac{d^{2} \vec{r}}{d t^{2}} = {} & -
      \frac{G M_{1}}{r^{3}} \vec{r} -
      \frac{G M_{2}}{\left \vert \vec{r}_{3} - \vec{r}_{2} \right \vert^{3}}
      \left ( \vec{r}_{3} - \vec{r}_{2} \right ) +
      \frac{G M_{2}}{\left \vert \vec{r}_{1} - \vec{r}_{2} \right \vert^{3}}
      \left ( \vec{r}_{1} - \vec{r}_{2} \right ) +
\notag \\
& + \beta \frac{G M_{1}}{r^{2}}
      \left [ \left ( 1 - \frac{\vec{v'} \cdot \vec{e}_{\text{R}}}{c} \right )
      \vec{e}_{\text{R}} - \frac{\vec{v'}}{c} \right ] +
      \frac{\eta}{\bar{Q'}_{\text{pr}}}
      \beta \frac{u'}{c} \frac{G M_{1}}{r^{2}} \left [
      \left ( 1 - \frac{\vec{v'} \cdot \vec{e}_{\text{R}}}{u'} \right )
      \vec{e}_{\text{R}} - \frac{\vec{v'}}{u'} \right ] ~.
\end{align}
The first three terms are caused by the Newton gravitation of the star and
the planet and the last two terms are the PR effect
\citep{icarus} and the radial stellar wind \citep{covsw}.
$\vec{r}_{i}$, $i$ $=$ 1, 2, 3, are, respectively, the positions vectors
of the star, planet and particle determined with respect to the center
of mass of the two heavier bodies. $\vec{v'}$ is the velocity
of the dust particle with respect to the star, $G$ is the gravitational
constant, $M_{1}$ is the mass of the star, and $M_{2}$ is the mass
of the planet. $\vec{r}$ $=$ $\vec{r}_{3} - \vec{r}_{1}$ is the
position vector of the particle with respect to the star,
$r$ $=$ $\left \vert \vec{r} \right \vert$ and $\vec{e}_{\text{R}}$ $=$
$\vec{r} / r$ is the unit vector directed from the star to the particle.
The parameter $\beta$ is defined as the ratio between the electromagnetic
radiation pressure force and the gravitational force between the star
and the particle at rest with respect to the star
\begin{equation}\label{beta}
\beta = \frac{L_{\star} A' \bar{Q'}_{\text{pr}}}{4 \pi c m G M_{1}} ~.
\end{equation}
Here, $L_{\star}$ is the stellar luminosity, $\bar{Q'}_{\text{pr}}$ is
the dimensionless efficiency factor for the radiation pressure
determined in the proper frame of the particle and averaged over
the stellar spectrum, $A'$ is the geometrical cross-section
of the spherical particle determined in the proper frame of the particle,
$c$ is the speed of light, and $m$ is the mass of the dust particle.
The parameter $\eta$ describes the magnitude of acceleration caused by
the radial stellar wind
\begin{equation}\label{eta}
\eta = \frac{4 \pi r^{2} u'}{L_{\star}}
\sum_{i = 1}^{N} n_{\text{sw}~i} m_{\text{sw}~i} c^{2} ~,
\end{equation}
where $m_{\text{sw}~i}$ and $n_{\text{sw}~i}$, $i$ $=$ 1 to $N$, are
the masses and concentrations of the stellar wind particles at a distance
$r$ from the star ($u'$ $=$ 450 km/s and $\eta$ $=$ 0.38 for the Sun,
\citealt{covsw}). To the given accuracy $\eta$ is the ratio of stellar
wind energy to stellar electromagnetic radiation energy, both radiated per
unit of time.

We assume that the mass of the particle is negligible with
respect to the mass of star and also with respect to the mass
of planet. Therefore, the Newton equations of motion of the two
heavier bodies, only under the action of their mutual gravitational
interaction, in the barycentric reference frame are
\begin{align}\label{Newton}
\frac{d^{2} \vec{r}_{1}}{d t^{2}} &= -
      \frac{G M_{2}}{\left \vert \vec{r}_{1} - \vec{r}_{2} \right \vert^{3}}
      \left ( \vec{r}_{1} - \vec{r}_{2} \right ) ~,
\notag \\
\frac{d^{2} \vec{r}_{2}}{d t^{2}} &= -
      \frac{G M_{1}}{\left \vert \vec{r}_{2} - \vec{r}_{1} \right \vert^{3}}
      \left ( \vec{r}_{2} - \vec{r}_{1} \right ) ~.
\end{align}
In order to obtain the equation of motion of the dust particle in
the barycentric reference frame we can use the following relation
\begin{equation}\label{relation}
\frac{d^{2} \vec{r}_{3}}{d t^{2}} =  \frac{d^{2} \vec{r}}{d t^{2}} +
\frac{d^{2} \vec{r}_{1}}{d t^{2}} ~.
\end{equation}
Eq. (\ref{relation}) with substituted Eq. (\ref{star}) and the first
of Eqs. (\ref{Newton}) yields
\begin{align}\label{bary}
      \frac{d \vec{v}}{d t} = {} & - \frac{G M_{1}}{r^{3}} \vec{r} -
      \frac{G M_{2}}{\left \vert \vec{r}_{3} - \vec{r}_{2} \right \vert^{3}}
      \left ( \vec{r}_{3} - \vec{r}_{2} \right ) +
      \beta \frac{G M_{1}}{r^{2}}
      \left [ \left ( 1 - \frac{\vec{v'} \cdot \vec{e}_{\text{R}}}{c} \right )
      \vec{e}_{\text{R}} - \frac{\vec{v'}}{c} \right ] +
\notag \\
& + \frac{\eta}{\bar{Q'}_{\text{pr}}}
      \beta \frac{u'}{c} \frac{G M_{1}}{r^{2}} \left [
      \left ( 1 - \frac{\vec{v'} \cdot \vec{e}_{\text{R}}}{u'} \right )
      \vec{e}_{\text{R}} - \frac{\vec{v'}}{u'} \right ] ~,
\end{align}
where $\vec{v}$ is the velocity of the dust particle in the barycentric
coordinate system. For stars similar to our Sun inequality
$( \eta / \bar{Q}'_{\text{pr}} ) ( u' / c )$ $\ll$ 1 holds
and Eq. (\ref{bary}) can be simplified
\begin{equation}\label{dust}
\frac{d \vec{v}}{d t} = -
\frac{G M_{1} \left ( 1 - \beta \right )}{r^{3}} \vec{r} -
\frac{G M_{2}}{\left \vert \vec{r}_{3} - \vec{r}_{2} \right \vert^{3}}
\left ( \vec{r}_{3} - \vec{r}_{2} \right ) -
\beta \frac{G M_{1}}{r^{2}}
\left ( 1 + \frac{\eta}{\bar{Q'}_{\text{pr}}} \right )
\left ( \frac{\vec{v'} \cdot \vec{e}_{\text{R}}}{c}
\vec{e}_{\text{R}} + \frac{\vec{v'}}{c} \right ) ~.
\end{equation}

In next step we will assume that we have found an equilibrium point for
the dust particle in a coordinate system rotating around the barycenter
with an angular velocity $n$ equal to the mean motion of the two-body
system. In the rotating coordinate system the dust particle located in
the equilibrium remains stationary. In the barycentric coordinate system
the dust particle located in the equilibrium point moves in a circular orbit
with a constant speed. Coordinates in the barycentric system will be denoted by
$x$, $y$ and $z$. The plane $x y$ in the chosen coordinate system
coincides with the orbital plane of the star and the planet. The planet
in the barycentric coordinate system moves counterclockwise when we look from
the positive $z$-axis. Let actual coordinates of the equilibrium point
in the barycentric system be $x$ and $y$. Then the velocity of the dust
particle in the barycentric coordinate system is
\begin{equation}\label{velod}
\vec{v} = \left ( -y n, ~x n, ~0 \right ) ~.
\end{equation}
Similarly for the star
\begin{equation}\label{velos}
\vec{v_{\text{S}}} = \left ( -y_{1} n, ~x_{1} n, ~0 \right ) ~.
\end{equation}
For the velocity of the dust particle in the reference frame associated
with the star we obtain
\begin{equation}\label{vprime}
\vec{v'} = \vec{v} - \vec{v_{\text{S}}} =
\left ( -\left ( y - y_{1} \right ) n, ~
\left ( x - x_{1} \right ) n, ~0 \right ) ~.
\end{equation}
For the unit vector $\vec{e}_{\text{R}}$ directed from the star to
the particle we obtain
\begin{equation}\label{eR}
\vec{e}_{\text{R}} = \frac{\vec{r}_{3} - \vec{r}_{1}}{r} = \frac{1}{r}
\left ( x - x_{1}, ~y - y_{1}, ~z \right ) ~.
\end{equation}
The scalar product in Eq. (\ref{dust}) on the basis of Eq. (\ref{vprime})
and Eq. (\ref{eR}) is
\begin{equation}\label{product}
\vec{v'} \cdot \vec{e}_{\text{R}} = 0 ~.
\end{equation}
Hence, only the term proportional to $\vec{v'}$ in the acceleration caused by
the non-gravitational effects remains and for the dust particle in
the equilibrium point we have (Eq. \ref{dust})
\begin{equation}\label{short}
\frac{d \vec{v}}{d t} = -
\frac{G M_{1} \left ( 1 - \beta \right )}{r^{3}} \vec{r} -
\frac{G M_{2}}{\left \vert \vec{r}_{3} - \vec{r}_{2} \right \vert^{3}}
\left ( \vec{r}_{3} - \vec{r}_{2} \right ) -
\beta \frac{G M_{1}}{c r^{2}}
\left ( 1 + \frac{\eta}{\bar{Q'}_{\text{pr}}} \right ) \vec{v'} ~.
\end{equation}
For the equilibrium points located out of the $xy$ plane we obtain from
the $z$-component of Eq. (\ref{short})
\begin{equation}\label{out}
\frac{G M_{1} \left ( 1 - \beta \right )}{r^{3}} +
\frac{G M_{2}}{\left \vert \vec{r}_{3} - \vec{r}_{2} \right \vert^{3}} = 0 ~.
\end{equation}
This equation does not have any solution for the dust particles with
$\beta$ $<$ 1 (for discussion of the out of plane equilibrium points with
$\beta$ $>$ 1 see \citealt{colombo}). In what follows we will consider only
the equilibrium points located in the $xy$ plane.

Introducing the transformation between the barycentric coordinate system
and the system rotating around the barycenter with the angular velocity $n$
\begin{align}\label{transformation}
x &= x_{\text{R}} \cos n t - y_{\text{R}} \sin n t ~,
\notag \\
y &= x_{\text{R}} \sin n t + y_{\text{R}} \cos n t
\end{align}
Eq. (\ref{short}) can be written in the following form
\begin{align}\label{derivatives}
\frac{d x_{\text{R}}^{2}}{d t^{2}} - 2 n \frac{d y_{\text{R}}}{d t} -
n^{2} x_{\text{R}} = {} & - \frac{G M_{1} \left ( 1 - \beta \right )}{r^{3}}
      \left ( x_{\text{R}} - x_{\text{1R}} \right ) -
      \frac{G M_{2}}{\left \vert \vec{r}_{3} - \vec{r}_{2} \right \vert^{3}}
      \left ( x_{\text{R}} - x_{\text{2R}} \right )
\notag \\
& + \beta \frac{G M_{1}}{c r^{2}}
      \left ( 1 + \frac{\eta}{\bar{Q'}_{\text{pr}}} \right ) n y_{\text{R}} ~,
\notag \\
\frac{d y_{\text{R}}^{2}}{d t^{2}} + 2 n \frac{d x_{\text{R}}}{d t} -
n^{2} y_{\text{R}} = {} & - \frac{G M_{1} \left ( 1 - \beta \right )}{r^{3}}
      y_{\text{R}} -
      \frac{G M_{2}}{\left \vert \vec{r}_{3} - \vec{r}_{2} \right \vert^{3}}
      y_{\text{R}}
\notag \\
& - \beta \frac{G M_{1}}{c r^{2}}
      \left ( 1 + \frac{\eta}{\bar{Q'}_{\text{pr}}} \right ) n
      \left ( x_{\text{R}} - x_{\text{1R}} \right ) ~,
\end{align}
where we have assumed that $y_{\text{1R}}$ $=$ $y_{\text{2R}}$ $=$ 0 holds
for the star and the planet. For an equilibrium point the time derivatives
of the coordinates in the rotating reference frame are zero and we obtain
\begin{align}\label{system}
n^{2} x_{\text{R}} &= \frac{G M_{1} \left ( 1 - \beta \right )}{r^{3}}
      \left ( x_{\text{R}} - x_{\text{1R}} \right ) +
      \frac{G M_{2}}{\left \vert \vec{r}_{3} - \vec{r}_{2} \right \vert^{3}}
      \left ( x_{\text{R}} - x_{\text{2R}} \right ) -
      \beta \frac{G M_{1}}{c r^{2}}
      \left ( 1 + \frac{\eta}{\bar{Q'}_{\text{pr}}} \right ) n y_{\text{R}} ~,
\notag \\
n^{2} y_{\text{R}} &= \frac{G M_{1} \left ( 1 - \beta \right )}{r^{3}}
      y_{\text{R}} +
      \frac{G M_{2}}{\left \vert \vec{r}_{3} - \vec{r}_{2} \right \vert^{3}}
      y_{\text{R}} + \beta \frac{G M_{1}}{c r^{2}}
      \left ( 1 + \frac{\eta}{\bar{Q'}_{\text{pr}}} \right ) n
      \left ( x_{\text{R}} - x_{\text{1R}} \right ) ~.
\end{align}

\subsection{Equilibrium points without velocity terms}
\label{sec:novelocity}

Equations (\ref{system}) without the radiation terms resulting from
the dependence of the acceleration in Eq. (\ref{dust}) on the relative
velocity of the dust particle with respect to the star is not real.
However, equilibrium points obtained from this equations can be used as first
approximation during determination of more real positions of the equilibrium
points. This is caused by the fact that from all terms in the acceleration
caused by the radiation the radial term not depending on the relative
velocity is dominant. The other terms are proportional to a ratio
of a component of the relative velocity and the speed of light.
By using this approximation we also assume that the terms resulting from
the dependence on the relative velocity of the dust particle with respect
to the star in Eqs. (\ref{system}) can be neglected in comparison
with the terms caused by the gravitational force of the planet. Positions
of the equilibrium points for the system given by Eq. (\ref{system}) without
velocity terms are determined by
\begin{align}\label{approx}
x_{\text{R}} n^{2} &= \frac{G M_{1} \left ( 1 - \beta \right )}{r^{3}}
      \left ( x_{\text{R}} - x_{\text{1R}} \right ) +
      \frac{G M_{2}}{\left \vert \vec{r}_{3} - \vec{r}_{2} \right \vert^{3}}
      \left ( x_{\text{R}} - x_{\text{2R}} \right ) ~,
\notag \\
y_{\text{R}} n^{2} &= \frac{G M_{1} \left ( 1 - \beta \right )}{r^{3}}
      y_{\text{R}} +
      \frac{G M_{2}}{\left \vert \vec{r}_{3} - \vec{r}_{2} \right \vert^{3}}
      y_{\text{R}} ~.
\end{align}
This system with $\beta$ $=$ 0 is the system which determines positions
of the Lagrangian points in the circular restricted
three-body problem. From the second equation we obtain two
possibilities one with $y_{\text{R}}$ $=$ 0 and one with
\begin{equation}\label{condition}
n^{2} = \frac{G M_{1} \left ( 1 - \beta \right )}{r^{3}} +
\frac{G M_{2}}{\left \vert \vec{r}_{3} - \vec{r}_{2} \right \vert^{3}} ~.
\end{equation}
From the second Kepler's law we have
\begin{equation}\label{Kepler}
n^{2} = \frac{G \left ( M_{1} + M_{2} \right )}{a_{\text{P}}^{3}} ~,
\end{equation}
where $a_{\text{P}}$ is the semimajor axis of the planet.

\subsubsection{Analogy with the Lagrangian points $L_{4}$ and $L_{5}$}
\label{sec:L45}

In the $x_{\text{P}} y_{\text{P}}$ plane only one distance
from the star and one from the planet fulfills the condition
Eq. (\ref{condition}) for a given $\beta$. The solution is
\begin{align}\label{L4L5}
r &= a_{\text{P}} \left ( 1 - \beta \right )^{1/3} ~,
\notag \\
\left \vert \vec{r}_{3} - \vec{r}_{2} \right \vert &= a_{\text{P}} ~.
\end{align}
Coordinates of the points analogous to the Lagrangian points $L_{4}$ and
$L_{5}$ are given by common points of the two circles given by
Eqs. (\ref{L4L5})
\begin{align}\label{coordinates}
x_{\text{L}} &= \frac{a_{\text{P}}
      \left [ \left ( 1 - \beta \right )^{2/3} - 1 \right ] +
      x_{\text{1R}} + x_{\text{2R}}}{2} ~,
\notag \\
y_{\text{L}}^{2} &= a_{\text{P}}^{2} \left \{ 1 - \left [ 1 -
      \frac{\left ( 1 - \beta \right )^{2/3}}{2} \right ]^{2} \right \} ~.
\end{align}

\begin{figure}[t]
\begin{center}
\includegraphics[height=0.31319097378751424591617069773315\textheight]{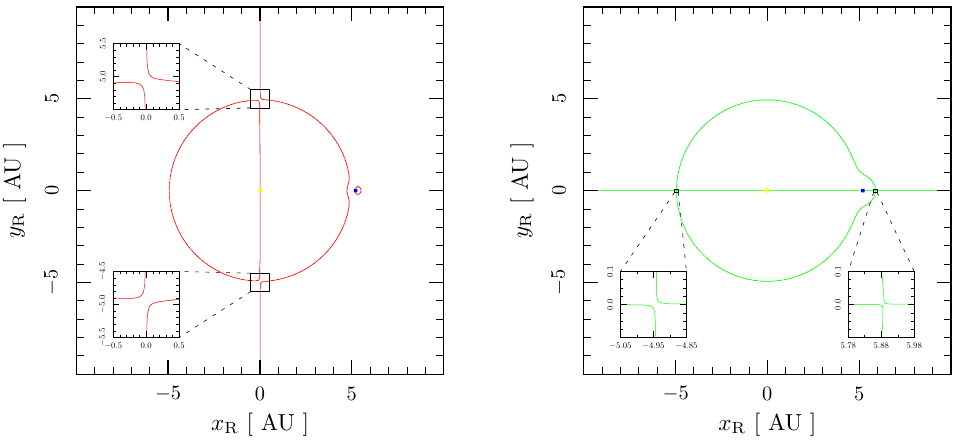}
\end{center}
\caption{Calculated points in the $x_{\text{R}} y_{\text{R}}$ plane
which fulfill the first (left) and the second (right) equation
in the system of equations given by Eqs. (\ref{system}).
The equations are solved for a dust particle with radius $R$ $=$ 4 $\mu$m,
density $\varrho$ $=$ 1 g/cm$^{3}$, and $\bar{Q}'_{\text{pr}}$ $=$ 1 in
the system with Jupiter in a circular orbit around the radiating Sun.}
\label{fig:system}
\end{figure}

\begin{figure}[t]
\begin{center}
\includegraphics[height=0.336\textheight]{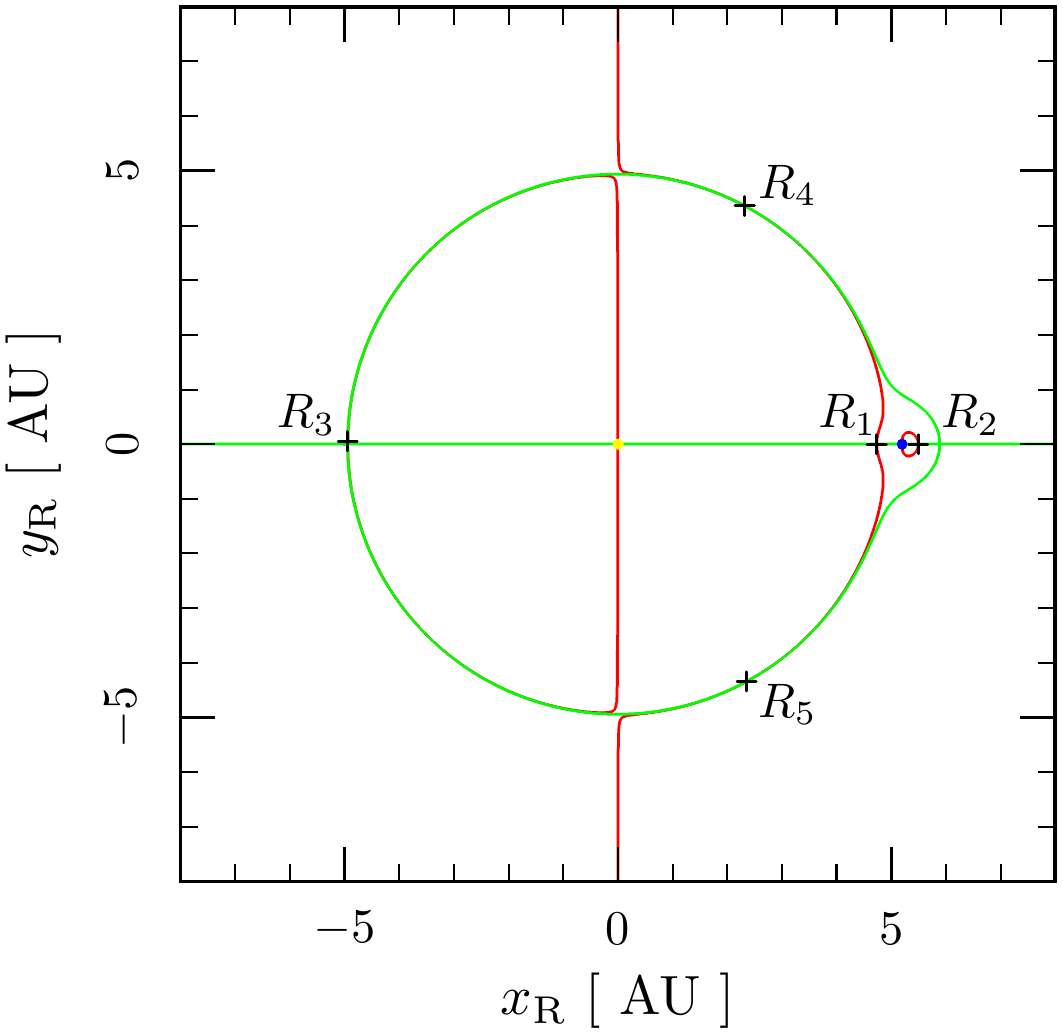}
\end{center}
\caption{The equilibrium points obtained from the solution
of Eqs. (\ref{system}) depicted in Fig. \ref{fig:system}. The system consists
of Jupiter in a circular orbit around the Sun and a dust particle with radius
$R$ $=$ 4 $\mu$m, mass density $\varrho$ $=$ 1 g/cm$^{3}$, and
$\bar{Q}'_{\text{pr}}$ $=$ 1. We must note that the point $R_{2}$ is inside
Jupiter's shadow for the real Jupiter.}
\label{fig:common}
\end{figure}

\subsubsection{Analogy with the Lagrangian points $L_{1}$, $L_{2}$ and
$L_{3}$}
\label{sec:L123}

The condition $y_{\text{R}}$ $=$ 0 for a given $\beta$ is fulfilled by
three different $x_{\text{R}}$ determined by the first equation in
Eqs. (\ref{approx}). For $x_{\text{1R}}$ and $x_{\text{2R}}$ we have from
the two-body problem
\begin{align}\label{ratio}
x_{\text{1R}} &= - \frac{M_{2} a_{\text{P}}}{M_{1} + M_{2}} ~,
\notag \\
x_{\text{2R}} - x_{\text{1R}} &= a_{\text{P}} ~.
\end{align}
When we look from the planet, then for the point before the star we obtain
\begin{equation}\label{L1}
\frac{M_{\text{2}}}{M_{\text{1}}} = \frac{\left ( 1 - \beta \right )
a_{\text{P}}^{3} - \left ( a_{\text{P}} - r_{\text{P}} \right )^{3}}
{\left ( a_{\text{P}} - r_{\text{P}} \right )^{2} \left ( a_{\text{P}}^{3} -
r_{\text{P}}^{3} \right )} r_{\text{P}}^{2} ~,
\end{equation}
where $r_{\text{P}}$ is the distance of the point from the planet. This point
is analogous to the Lagrangian point $L_{1}$. The equilibrium point
analogous to the Lagrangian point $L_{2}$ is located in the opposite
direction from the planet. From the first equation in Eqs. (\ref{approx})
we have
\begin{equation}\label{L2}
\frac{M_{\text{2}}}{M_{\text{1}}} = \frac{\left ( 1 - \beta \right )
a_{\text{P}}^{3} - \left ( a_{\text{P}} + r_{\text{P}} \right )^{3}}
{\left ( a_{\text{P}} + r_{\text{P}} \right )^{2} \left ( r_{\text{P}}^{3} -
a_{\text{P}}^{3} \right )} r_{\text{P}}^{2} ~.
\end{equation}
Finally, the point analogous to the Lagrangian point $L_{3}$ is located
behind the star with
\begin{equation}\label{L3}
\frac{M_{\text{2}}}{M_{\text{1}}} = \frac{\left ( 1 - \beta \right )
a_{\text{P}}^{3} + \left ( a_{\text{P}} - r_{\text{P}} \right )^{3}}
{\left ( a_{\text{P}} - r_{\text{P}} \right )^{2} \left ( r_{\text{P}}^{3} -
a_{\text{P}}^{3} \right )} r_{\text{P}}^{2} ~.
\end{equation}
Coordinates of the points are
$x_{\text{L}}$ $=$ $x_{\text{2R}}$ $-$ $r_{\text{P}}$,
$x_{\text{L}}$ $=$ $x_{\text{2R}}$ $+$ $r_{\text{P}}$, and
$x_{\text{L}}$ $=$ $x_{\text{2R}}$ $-$ $r_{\text{P}}$
with $y_{\text{L}}$ $=$ 0, where $r_{\text{P}}$ has to be determined
by Eq. (\ref{L1}), (\ref{L2}), and (\ref{L3}), respectively.

\subsection{Equilibrium points with velocity terms}
\label{sec:velocity}

If are the velocity terms in the acceleration caused by the radiation
taken into account, then Eqs. (\ref{system}) determine positions
of the equilibrium points exactly. One solution of Eqs. (\ref{system})
is shown in Fig. \ref{fig:system}. We assumed that the system
consists of Jupiter in a circular orbit around the Sun and
a dust particle with radius $R$ $=$ 4 $\mu$m, mass density
$\varrho$ $=$ 1 g/cm$^{3}$, and $\bar{Q}'_{\text{pr}}$ $=$ 1. The left part
of the figure corresponds to the first equation in Eqs. (\ref{system}) and
the right part of the figure corresponds to the second equation in
Eqs. (\ref{system}). The solution of the first equation is formed by three
separated curves which do not intersect each other. The solution of the second
equation is formed by a single curve without intersection. Theoretically,
intersections of the curve corresponding to the solution of the second
equation exist for no radiation case ($\beta$ $=$ 0) because $y_{\text{R}}$
$=$ 0 is always one solution. The equilibrium points can be obtained as
common points of the curves were both equations in Eqs. (\ref{system}) hold
at once. The equilibrium points determined by the common points of the curves
in Fig. \ref{fig:system} are depicted in Fig. \ref{fig:common}.
Five equilibrium points exist in this case. This numerical method
always allows the determination of the positions of the equilibrium points.

Figures \ref{fig:system} and \ref{fig:common} show that the solutions
of the system of equations in Eqs. (\ref{system}) have relative complicated
behavior. We must have in mind that these figures hold only for a single value
of $\beta$. In order to find the positions of the equilibrium points
analytically and overcome the relatively complicated behavior of solutions
of Eqs. (\ref{system}) we developed an analytical method which will be now
presented. Suppose that the terms resulting from the dependence of acceleration
of the dust particle on the relative velocity are small in comparison
with all other terms in Eqs. (\ref{system}). In this case the positions
of the equilibrium points will be approximately determined by the theory
presented in Section \ref{sec:novelocity}. The last two terms in
Eqs. (\ref{system}) will create only a shift from the positions analogous
to the Lagrangian points. We will denote this shift as
$\Delta x_{\text{R}}$ and $\Delta y_{\text{R}}$. The position analogous
to the Lagrangian points will denoted by $x_{\text{L}}$ and $y_{\text{L}}$.
Using a Taylor series for functions with two variables to first order
\begin{equation}\label{taylor}
f \left ( x_{\text{R}}, y_{\text{R}} \right ) \approx
f \left ( x_{\text{L}}, y_{\text{L}} \right ) +
\left ( \frac{\partial f}{\partial x_{\text{R}}} \right )
_{x_{\text{L}}, y_{\text{L}}}
\left ( x_{\text{R}} - x_{\text{L}} \right ) +
\left ( \frac{\partial f}{\partial y_{\text{R}}} \right )
_{x_{\text{L}}, y_{\text{L}}}
\left ( y_{\text{R}} - y_{\text{L}} \right ) ~.
\end{equation}
we can approximate the expressions with the distances in Eqs. (\ref{system})
\begin{align}\label{distances}
\frac{1}{\left \vert \vec{r}_{3} - \vec{r}_{1} \right \vert^{3}} &\approx
      \frac{1}{r_{\text{1L}}^{3}} +
      \frac{3 \left ( x_{\text{1R}} - x_{\text{L}} \right )}{r_{\text{1L}}^{5}}
      \Delta x_{\text{R}} -
      \frac{3 y_{\text{L}}}{r_{\text{1L}}^{5}} \Delta y_{\text{R}} ~,
\notag \\
\frac{1}{\left \vert \vec{r}_{3} - \vec{r}_{1} \right \vert^{2}} &\approx
      \frac{1}{r_{\text{1L}}^{2}} +
      \frac{2 \left ( x_{\text{1R}} - x_{\text{L}} \right )}{r_{\text{1L}}^{4}}
      \Delta x_{\text{R}} -
      \frac{2 y_{\text{L}}}{r_{\text{1L}}^{4}} \Delta y_{\text{R}} ~,
\notag \\
\frac{1}{\left \vert \vec{r}_{3} - \vec{r}_{2} \right \vert^{3}} &\approx
      \frac{1}{r_{\text{2L}}^{3}} +
      \frac{3 \left ( x_{\text{2R}} - x_{\text{L}} \right )}{r_{\text{2L}}^{5}}
      \Delta x_{\text{R}} -
      \frac{3 y_{\text{L}}}{r_{\text{2L}}^{5}} \Delta y_{\text{R}} ~,
\end{align}
where
\begin{align}\label{rLs}
r_{\text{1L}} &= \sqrt{\left ( x_{\text{1R}} - x_{\text{L}} \right )^{2} +
      y_{\text{L}}^{2}} ~,
\notag \\
r_{\text{2L}} &= \sqrt{\left ( x_{\text{2R}} - x_{\text{L}} \right )^{2} +
      y_{\text{L}}^{2}} ~.
\end{align}
Substitution of Eqs. (\ref{distances}) into Eqs. (\ref{system}) leads to
the following system of equations with two unknowns $\Delta x_{\text{R}}$
and $\Delta y_{\text{R}}$
\begin{align}\label{unknowns}
a_{11} \Delta x_{\text{R}} + a_{12} \Delta y_{\text{R}} + b_{1} = 0 ~,
\notag \\
a_{21} \Delta x_{\text{R}} + a_{22} \Delta y_{\text{R}} + b_{2} = 0 ~,
\end{align}
where
\begin{align}\label{coefficients}
a_{11} = {} & 2 n^{2} + \frac{G M_{1} \left ( 1 - \beta \right )}
      {r_{\text{1L}}^{3}} + \frac{G M_{2}}{r_{\text{2L}}^{3}} - a_{22} ~,
\notag \\
a_{12} = {} & \frac{3 G M_{1} \left ( 1 - \beta \right )}{r_{\text{1L}}^{5}}
      \left ( x_{\text{L}} - x_{\text{1R}} \right ) y_{\text{L}} +
      \frac{3 G M_{2}}{r_{\text{2L}}^{5}}
      \left ( x_{\text{L}} - x_{\text{2R}} \right ) y_{\text{L}}
\notag \\
& + \frac{\beta G M_{1}}{c}
      \left ( 1 + \frac{\eta}{\bar{Q'}_{\text{pr}}} \right ) n
      \left ( \frac{1}{r_{\text{1L}}^{2}} -
      \frac{2 y_{\text{L}}^{2}}{r_{\text{1L}}^{4}} \right ) ~,
\notag \\
a_{21} = {} & a_{12} ~,
\notag \\
a_{22} = {} & n^{2} - \frac{G M_{1} \left ( 1 - \beta \right )}
      {r_{\text{1L}}^{3}} - \frac{G M_{2}}{r_{\text{2L}}^{3}} +
      \frac{3 G M_{1} \left ( 1 - \beta \right )}{r_{\text{1L}}^{5}}
      y_{\text{L}}^{2}
\notag \\
& + \frac{3 G M_{2}}{r_{\text{2L}}^{5}} y_{\text{L}}^{2} +
      \frac{2 \beta G M_{1}}{c}
      \left ( 1 + \frac{\eta}{\bar{Q'}_{\text{pr}}} \right ) n
      \frac{\left ( x_{\text{L}} - x_{\text{1R}} \right ) y_{\text{L}}}
      {r_{\text{1L}}^{4}} ~,
\notag \\
b_{1} = {} & \frac{\beta G M_{1}}{c}
      \left ( 1 + \frac{\eta}{\bar{Q'}_{\text{pr}}} \right ) n
      \frac{y_{\text{L}}}{r_{\text{1L}}^{2}} ~,
\notag \\
b_{2} = {} & - \frac{\beta G M_{1}}{c}
      \left ( 1 + \frac{\eta}{\bar{Q'}_{\text{pr}}} \right ) n
      \frac{x_{\text{L}} - x_{\text{1R}}}{r_{\text{1L}}^{2}} ~.
\end{align}
Eqs. (\ref{unknowns}) and (\ref{coefficients}) enable to determine
unknown shift from the points analogous to the Lagrangian points.
The determination of the shift from the points analogous
to $L_{4}$ and $L_{5}$ using Eqs. (\ref{coefficients}) can be
simplified with identity
\begin{equation}\label{identity}
n^{2} = \frac{G M_{1} \left ( 1 - \beta \right )}
{r_{\text{1L}}^{3}} + \frac{G M_{2}}{r_{\text{2L}}^{3}}
\end{equation}
and the substitution $y_{\text{L}}$ $=$ 0 simplify Eqs. (\ref{coefficients})
for the shift from the points analogous to $L_{1}$, $L_{2}$ and $L_{3}$.

\begin{figure}[t]
\begin{center}
\includegraphics[height=0.336\textheight]{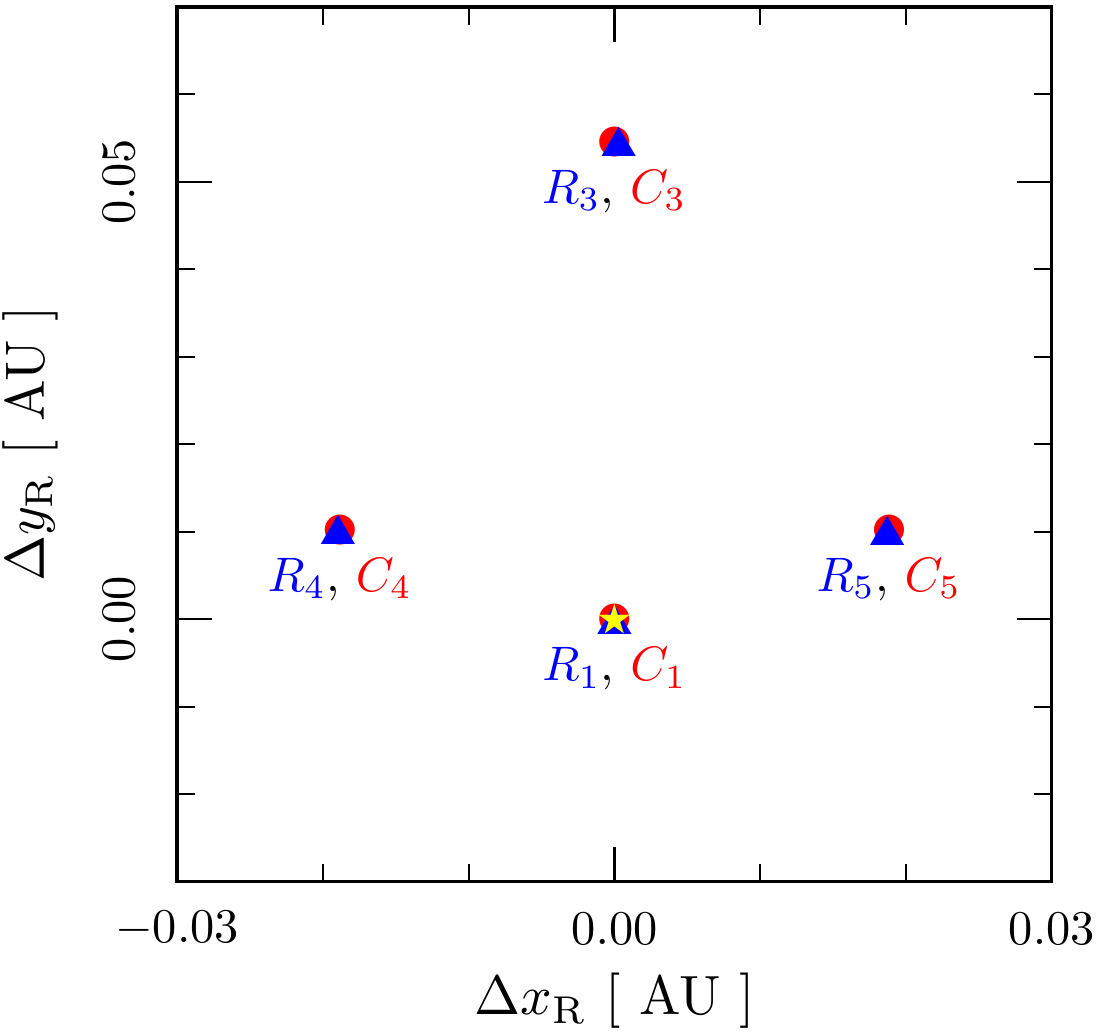}
\end{center}
\caption{Calculated ($C$) and real ($R$) shifts of the positions of dust
particle from the points analogous to the Lagrangian points in the CR3BP
with radiation. The shifts are determined for the system solved in
Fig. \ref{fig:system} and Fig. \ref{fig:common}. The circles are used for
the calculated positions, the triangles are used for the real positions
and the star denotes the position of the points analogous to the Lagrangian
points.}
\label{fig:delta}
\end{figure}

\begin{figure}[t]
\begin{center}
\includegraphics[height=0.33658156128576474743906746732585\textheight]{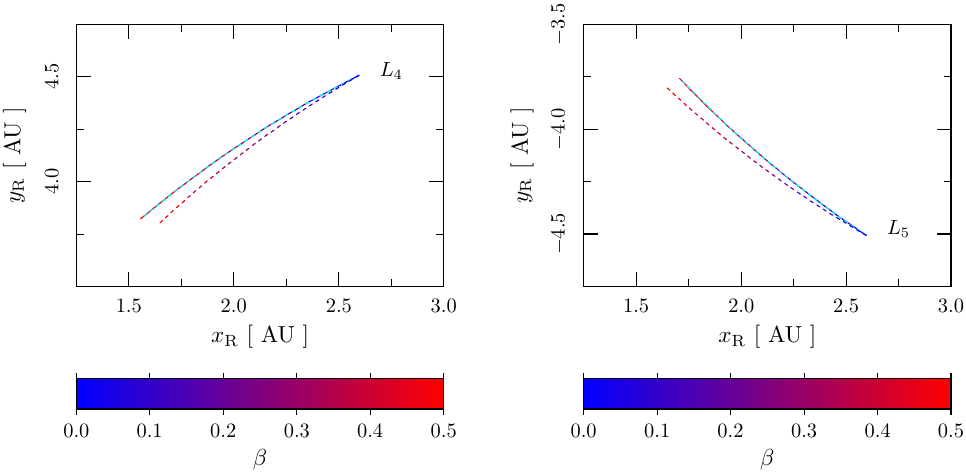}
\end{center}
\caption{The positions of real equilibrium points (solid line) compared with
the positions of equilibrium points calculated from Eqs. (\ref{unknowns})
(azure dashed line) and the positions of points analogous to the Lagrangian
points (dotted line) for Jupiter in a circular orbit around the Sun and dust
particles with $\beta$ $\in$ [0, 0.5] and $\bar{Q'}_{\text{pr}}$ $=$ 1.
Only the stable equilibrium points resulting from the points analogous
to the Lagrangian points $L_{4}$ and $L_{5}$ are depicted.}
\label{fig:stable}
\end{figure}

Figure \ref{fig:delta} shows the shifts of equilibrium points from
positions analogous to the Lagrangian points determined using system
of equation Eqs. (\ref{unknowns}) (circles) and the real shifts determined
using the numerical solution of Eqs. (\ref{system}) (triangles).
The parameters of system are the same as for Fig. \ref{fig:system} and
Fig. \ref{fig:common}. As can be seen the positions of equilibrium points
differ from the positions of points analogous to $L_{3}$, $L_{4}$ and $L_{5}$.
The position of equilibrium point derived from the point analogous
to $L_{1}$ is determined mainly by the gravitational influence
of the planet in this case and thus its position is not significantly
affected by the radiation terms depending on the relative velocity.
The point analogous to $L_{2}$ is inside Jupiter's shadow and is not shown.

Figure \ref{fig:stable} shows the positions of the stable equilibrium
points resulting from the points analogous to the Lagrangian
points $L_{4}$ and $L_{5}$ in the system consisting of Jupiter
in a circular orbit around the Sun, and the dust particles with
$\beta$ $\in$ [0, 0.5] and $\bar{Q'}_{\text{pr}}$ $=$ 1.
The positions of equilibrium points determined
from the numerical solution of Eqs. (\ref{system}) (solid line) are
compared with the positions of equilibrium points determined from
Eqs. (\ref{unknowns}) (azure dashed line). The positions of points
analogous to the Lagrangian points $L_{4}$ and $L_{5}$ are also shown
(dotted line). In Fig. \ref{fig:stable} we can see that the theory leading
to the system of equations represented by Eqs. (\ref{unknowns}) is
in an excellent agreement with the numerical solution of Eqs. (\ref{system})
for the considered system.

\begin{figure}[t]
\begin{center}
\includegraphics[height=0.43090555135783812613893950842258\textheight]{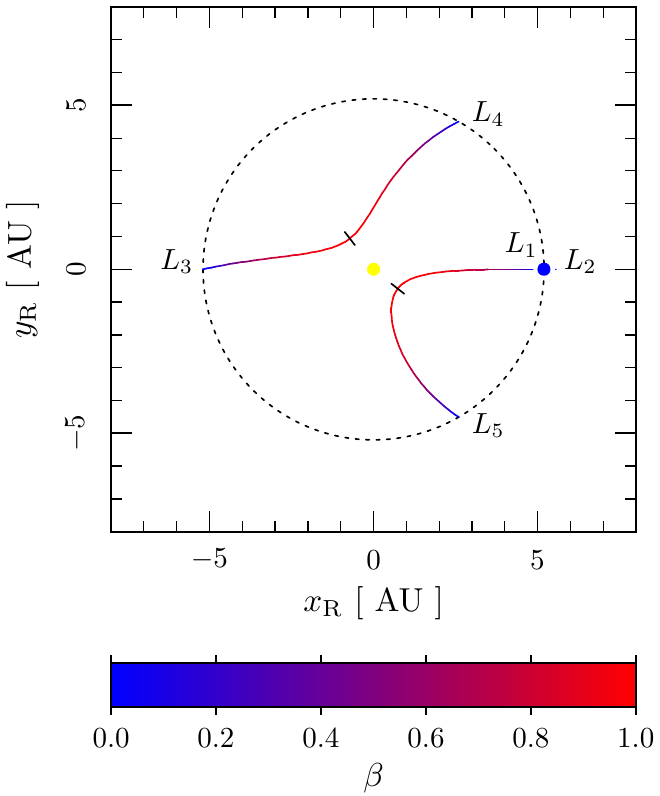}
\end{center}
\caption{Complete set of the equilibrium points in the circular restricted
three-body problem with radiation. Jupiter in a circular orbit around
the Sun is taken into account. The Lagrangian points $L_{1}$ and $L_{5}$
create an equilibrium branch. Similarly as $L_{3}$ and $L_{4}$.
Black lines perpendicular on the branches separate stable and unstable
parts of the branches. Only the Lagrangian point $L_{2}$ is depicted behind
Jupiter since the other equilibrium points resulting from $L_{2}$ are
inside Jupiter's shadow.}
\label{fig:jupiter}
\end{figure}

Figure \ref{fig:jupiter} shows complete set of the equilibrium
points for Jupiter and the Sun. It is purely theoretical case
in the CR3BP with included radiation. In Fig. \ref{fig:jupiter}
we can see that as $\beta$ increases from zero the equilibrium
points emerge from the Lagrangian points. The Lagrangian points
$L_{1}$ and $L_{5}$ are connected with a curve determined by
locations of the equilibrium points. We will call such curves
equilibrium branches or simply branches. Similarly the Lagrangian
points $L_{3}$ and $L_{4}$ create an equilibrium branch.

In \citet{murray} the first branch is obtained between
$L_{2}$ and $L_{5}$ and the second branch is obtained between
$L_{3}$ and $L_{4}$. A method used in \citet{murray} does not
eliminate dependence on $\beta$ in a resulting equation equivalent
to Eq. (99) in \citet{murray} for the complete acceleration caused
by the PR effect and the radial solar wind. If this method
would be used for Eqs. (\ref{system}), then the dependence
on $\eta / \bar{Q'}_{\text{pr}}$ would be eliminated.
This leads to equilibrium points obtained for a given $\beta$ and all
possible $\eta / \bar{Q'}_{\text{pr}}$. This is not appropriate
for a given star. Dependence on $\beta$ remains in the resulting
equation due to the radiation term which does not depend
on the relative velocity in Eq. (\ref{dust}). Moreover, the drag
term considered in \citet{murray} depends on the velocity of the dust
particle with respect to the center of mass of the two heavier bodies
(compare Eqs. 93 and 121 in \citealt{murray}) and the PR effect
give dependence on the relative velocity of the dust
particle with respect to the star (Eq. \ref{short}).
The dependence on the velocity of the dust particle with respect
to the center of mass of the two-body system in the PR effect
and the radial solar wind was also used in \citet{LZJ}.
They numerically obtained positions for the equilibrium points for
$\beta$ $\in$ \{0, 0.1, 0.2, ..., 0.9\}. Their Fig. 2 at least
qualitatively corresponds with the results presented
in Fig. \ref{fig:jupiter}. However, from Fig. 2 in \citet{LZJ} can
not be determined which Lagrangian points are connected with
equilibrium branches.

The stability of the equilibrium points can be verified directly
from the numerical solution of the equation of motion (Eq. \ref{dust}).
A point which separates branches into stable and unstable
parts exists on both branches in Fig. \ref{fig:jupiter}.
The location of this separation point is depicted with
a black line perpendicular on both branches. The separation point for
$L_{1}$--$L_{5}$ branch is closer then 1 AU from the Sun. The separation
point is the only equilibrium point on the $L_{1}$--$L_{5}$ branch for
$\beta$ $\approx$ 0.9935. For larger values of $\beta$ no equilibrium point
exists on the $L_{1}$--$L_{5}$ branch. The part of the branch
$L_{1}$--$L_{5}$ from the Lagrangian point $L_{5}$ to the separation
point is stable and the remaining part of the branch
is unstable. Similarly for the branch $L_{3}$--$L_{4}$.
The separation point on the $L_{3}$--$L_{4}$ branch is theoretically
obtained for a dust particle with $\beta$ $\approx$ 0.9880. If the particle
is initially near its equilibrium point on the stable part of a branch, then
the particle will librate around the equilibrium point. The particle near
its equilibrium point on the unstable part of the branch does not librate and
its distance from the equilibrium point gradually increases. The libration
around an equilibrium point on the stable part of branch is characterized
with small increase of the maximal distance from the equilibrium point
(libration amplitude) after one libration period. Therefore, the particles
librating around the equilibrium points on the stable part of branch
cannot remain in the libration and need to be replenished. The increase
of libration amplitude (instability of the equilibrium points) can be derived
analytically (see the comment after Eq. \ref{distance}). As the particles
librate time variations in the solar wind can also change the libration
amplitude. A particle with $\beta$ $<$ 1 can move in a bound orbit after
an ejection from a parent body. A particle with $\beta$ $>$ 1
escapes from the Solar system as $\beta$-meteoroid
\citep{meteoroids}. We must note that for the particles with
large $\beta$ (submicrometer particles) the Lorentz force
cannot be always neglected in comparison with the PR effect and
the radial solar wind \citep{dohnanyi,leinert,dermott}.
If the dust particle is already in the bound orbit,
then the PR effect and the radial solar wind
decrease the semimajor axis and eccentricity of the particle's orbit
\citep{ww}. Hence, the particle can get closer to circular orbits from
larger values of semimajor axes. This processes can take place after
the ejection of the dust particles from asteroids after mutual collisions
or from comets during their approach to the Sun. The dust particles
spiralling toward the Sun can get to the vicinity of the equilibrium
points.

\begin{figure}[t]
\begin{center}
\includegraphics[height=0.50429121294279745766478510156006\textheight]{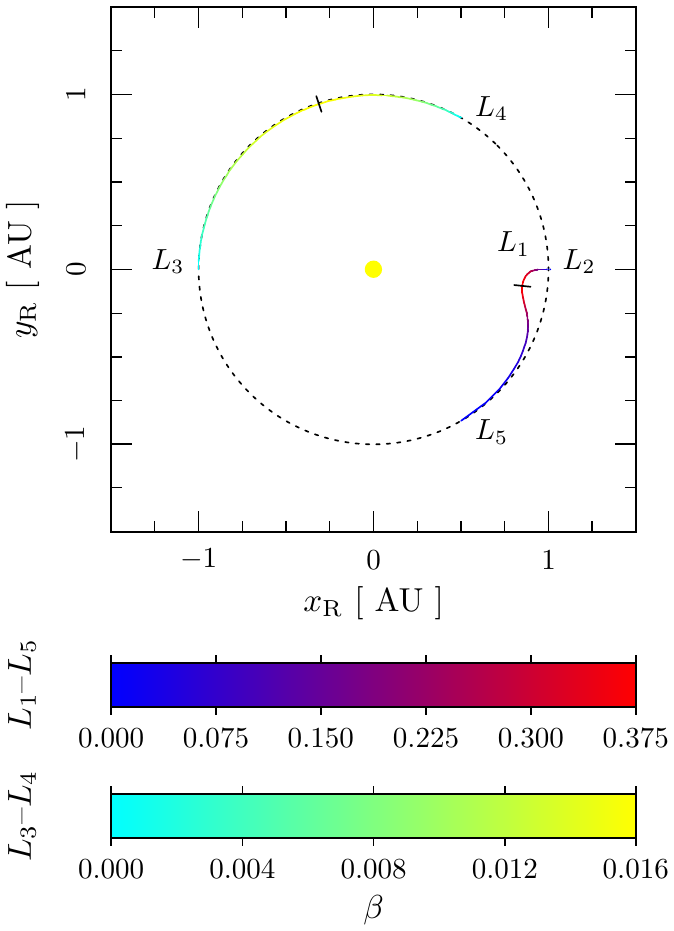}
\end{center}
\caption{Positions of the equilibrium points of dust particles
determined for the radiating Sun and the Earth in a circular orbit
in the reference frame rotating with the Earth. The Lagrangian points
$L_{1}$, $L_{3}$, $L_{4}$ $L_{5}$ are on the ends of equilibrium branches
$L_{1}$--$L_{5}$ and $L_{3}$--$L_{4}$. The equilibrium points on the branch
$L_{1}$--$L_{5}$ can be obtained for significantly smaller particles
(larger $\beta$) than the smallest particle in the equilibrium point
on the branch $L_{3}$--$L_{4}$. The stable part of branch
$L_{1}$--$L_{5}$ gets close to the Earth in the trailing direction.}
\label{fig:earth}
\end{figure}

Figure \ref{fig:earth} depicts locations of the equilibrium
points for the Earth in the CR3BP with radiation.
The black lines perpendicular to the equilibrium branches
split the branches into stable and unstable parts in
Fig. \ref{fig:earth}. Similarly to Fig. \ref{fig:jupiter}.
Only the Lagrangian point $L_{2}$ is shown behind the Earth.
The others equilibrium points resulting from the equilibrium near $L_{2}$
are partially in shadow cast by the Earth in the sunlight. The Lagrangian
point $L_{2}$ is 0.0100 AU behind the Earth and full shadow of the real
Earth extend to $\sim$0.0093 AU. These equilibrium points are also engulfed
in the magnetosphere of the Earth. The magnetotail of the Earth can exceed
to 0.067 AU.

\begin{figure}[t]
\begin{center}
\includegraphics[height=0.42620134228187919463087248322125\textheight]{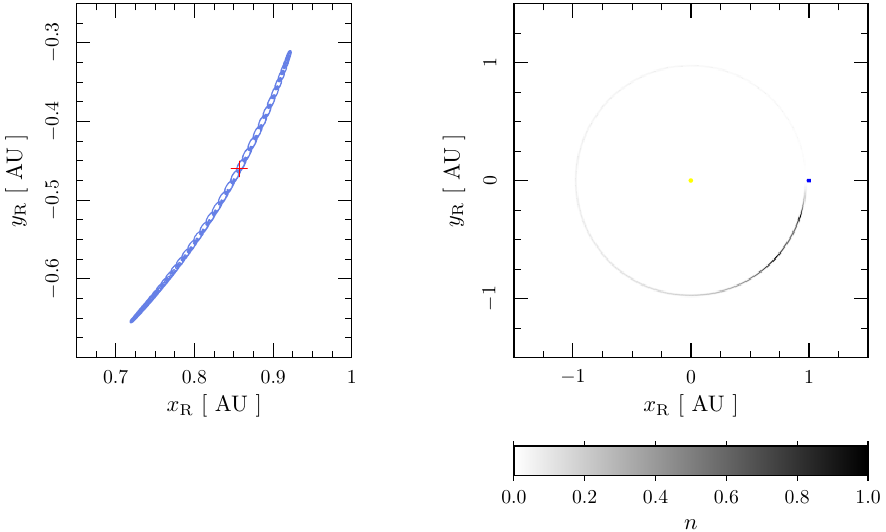}
\end{center}
\caption{Left: Libration of a dust particle with $\beta$ $=$ 0.08 and
$\bar{Q'}_{\text{pr}}$ $=$ 1 around its equilibrium point on the stable
part of the branch $L_{1}$--$L_{5}$. The position of the equilibrium
is depicted with a red cross. Right: A ring formed by particles with the same
properties and the initial conditions as the particle in the left plot.
Number density $n$ of the chosen particle decreases with increasing
circumferential distance from the initial libration depicted in
the left plot.}
\label{fig:libration}
\end{figure}

A circumsolar dust ring close to Earth's orbit was observed by satellites
{\it IRAS} \citep{IRAS} and {\it COBE} \citep{COBE}. The dust particles in
this ring can comprise particles moving close to the equilibrium points
of the CR3BP with radiation. If we take into account these particles, then
the cloud trailing the Earth observed with the {\it Spitzer Space Telescope}
\citep{reach} can be easily explained. The particles
forming the cloud librate around of the equilibrium points
on the stable part of branch $L_{1}$--$L_{5}$. Fig. \ref{fig:earth} shows
that the stable part of branch $L_{1}$--$L_{5}$ gets close to the Earth in
the trailing direction. The position of the cloud trailing the Earth
observed with the {\it Spitzer Space Telescope} is in an accordance with
consequences of used accelerations for the Poynting--Robertson effect
and the radial solar wind. The left plot in Fig. \ref{fig:libration}
shows a dust particle with $\beta$ $=$ 0.08 and $\bar{Q'}_{\text{pr}}$ $=$ 1
during 200 years of the libration around its equilibrium point
on the stable part of branch $L_{1}$--$L_{5}$. If this particle
would be continually replenished with the same initial
conditions, then all dust particles would form a ring near
the Earth's orbit depicted in the right plot of Fig. \ref{fig:libration}.
As the libration amplitude increases the particles spread from the equilibrium
point around the orbit and form the ring. Number density $n$
of the particles in the ring decreases with increasing circumferential
distance from the initial libration closest to the equilibrium point.
During conjunctions of a dust particle with Jupiter the gravitation
of Jupiter can be comparable with acceleration caused by the solar radiation
in the vicinity of the Earth's orbit (see Appendix A2 in \citealt{colombo}).
We have found using the numerical solution of the equation of motion
of the dust particle in a planar restricted four-body problem
with radiation that the libration of the dust particle close to
the Earth trailing direction is not affected by the gravitation
of Jupiter.

The mean motion resonances (MMRs) occur when orbital periods
of the particle and the Earth are in a ratio of natural numbers.
If the dust particle is captured in an MMR, then variations
of the semimajor axis are balanced by the resonant
interaction with the planet's gravitational field.
The capture is only temporary \citep{gomes}. The semimajor
axis of the particle's orbit in an MMR can be calculated using
the third Kepler laws for the particle and the Earth. The result is
\begin{align}\label{axis}
a_{\beta} &= a_{\text{P}} \left ( 1 - \beta \right )^{1/3}
\left ( \frac{T}{T_{\text{P}}} \right )^{2/3}
\left ( \frac{M_{1}}{M_{1} + M_{2}} \right )^{1/3} ~,
\notag \\
\frac{T}{T_{\text{P}}} &= \frac{p}{q} ~,
\end{align}
where $T$ is the orbital period of the dust particle,
$T_{\text{P}}$ is the orbital period of the Earth and $p$ and $q$
are two natural numbers. $p$ $>$ $q$ holds for the particle
captured in an exterior MMR and $p$ $<$ $q$ holds for
the capture in an interior MMR. The dust particles
captured in the MMRs with the Earth undergo
changes of their orbits caused by the solar radiation.
From typical dust particles ($\beta$ $\ll$ 1)
close to the Earth's orbit can stay only the particles
captured in MMRs with $p / q$ $\approx$ 1. The particles cannot
be captured in the MMRs with $p / q$ $\approx$ 1
such a long time as particles captured in the MMRs
given by a ratio of two small natural numbers due to
a lack of encounters with the Earth. This holds for
both the exterior and interior MMRs. Therefore,
the most suitable resonance for the formation
of the circumsolar dust ring close to Earth's
orbit is a special case, the resonance with $p / q$ $=$ 1/1 $=$ 1.
This resonance has also long capture time in comparison
with the capture times of resonances $p / q$ $\approx$ 1.
Since the gravitation of the Sun is dominant for
the dust particles in the equilibrium points approximately hold
\begin{equation}\label{distance}
r \approx a_{\text{P}} \left ( 1 - \beta \right )^{1/3} ~,
\end{equation}
where $r$ is the heliocentric distance. Similar equation can be obtained
from Eq. (\ref{axis}) if we consider the dust particle in the 1/1 resonance
in a circular orbit. The ring formed by the identical
particles depicted in Fig. \ref{fig:libration} represents
one possible behavior of the dust particles captured into the 1/1 resonance
with the Earth. The spherical dust particles captured in the 1/1 resonance
in the planar CR3BP under the action of the PR effect and the radial stellar
wind have an evolution of eccentricity which asymptotically decreases to
a zero eccentricity \citep{evo,mmrflow}. This feature enables using
an application of adiabatic invariant theory presented in \citet{gomes}
to prove that during the libration around an equilibrium point
on the stable part of an equilibrium branch the libration amplitude must
always increase. However, we can confirm result of \citet{LZJ} that
it is impossible for a dust particle to drift toward and get trapped
in the 1/1 resonance when the dust particle is approaching under
the PR effect and radial stellar wind. We found that such a capture
is possible after close encounter with the planet.

Figures \ref{fig:stable}, \ref{fig:jupiter} and \ref{fig:earth}
depicts that the equilibrium points are in interplanetary space
distributed in space according to the properties of dust particles.
This leads to applicable consequences. If the size and the mass
of the dust particles could be measured during an interplanetary flight
of a dust analyzing spacecraft, then parameter $\beta$ can be obtained for
the measured particles (with the assumption $\bar{Q'}_{\text{pr}}$ $=$ 1).
For a spacecraft trajectory going through the equilibrium branches
detection of the particles with the equilibrium properties should be
more frequent. Such a measurement could test the applicability
of the acceleration caused by the PR-effect and radial solar wind for
the description of motion of the real interplanetary dust particles.

Interstellar medium atoms penetrate into the Solar System due to
relative motion of the Solar System with respect to the interstellar
medium. These approaching atoms form an interstellar gas flow.
In the outer parts of the Solar system the acceleration of the dust
particle caused by the interstellar gas flow cannot be neglected with
respect to the accelerations caused by the PR effect and the radial
solar wind \citep*{flow,dyncd}. The interstellar gas flow significantly
affects also the orbital evolution of the dust particles in the mean motion
resonances \citep{mmrflow,stab}. Since the interstellar gas flow
is monodirectional its implementation in the rotating reference frame
in order to determine the positions of the equilibrium points is
impossible. Hence, in the outer parts of the Solar system
the equilibrium discussed in this paper cannot be obtained.

\section{Conclusion}
\label{sec:conclusion}

Equilibrium points exist for dust particles in the circular restricted
three-body problem with radiation. Positions of the equilibrium points
can be calculated with presented analytical method in the case when
the terms resulting from the dependence of the acceleration of dust particle
on the relative velocity between the particle and the star are small
in comparison with all other terms in the acceleration.
This assumption is not valid for the Earth due to its small mass.
The Lagrangian points $L_{1}$ and $L_{5}$ are connected into an equilibrium
branch for the dust particles. Similarly for the Lagrangian points $L_{3}$
and $L_{4}$. The equilibrium points of the Sun and Jupiter are located far
from Jupiter's orbit. The equilibrium points for the Earth explain
the cloud of dust particles trailing the Earth. The dust particles moving
in the equilibrium points are distributed in the interplanetary space
according to their properties. This should have interesting applications
in tests of the accelerations acting on the real interplanetary dust
particles.

\section*{Acknowledgement}

I would like to thank the anonymous reviewer of this paper for his useful
comments and suggestions.

\end{document}